\documentclass{edm_article}

\usepackage{xcolor}
\usepackage{graphicx}
\usepackage{subcaption}
\usepackage{tikz, float}
\usepackage{xparse}
\usepackage{xcolor}

\definecolor{businessblue}{HTML}{003d9e}
\definecolor{businessblue2}{HTML}{0077d3}
\definecolor{businessblue3}{HTML}{00aada}
\definecolor{businessblue4}{HTML}{00d8b7}
\definecolor{businessblue5}{HTML}{75ff85}
\definecolor{psychpopi}{HTML}{7100a6}
\definecolor{bizpsych}{HTML}{3561db}
\definecolor{psychpopii}{HTML}{9cedff}
\definecolor{sleep}{HTML}{0096f3}
\definecolor{newpsych}{HTML}{3dc4fb}
\definecolor{engi1}{HTML}{1778ff}
\definecolor{engi2}{HTML}{1db9ff}
\definecolor{engi3}{HTML}{00d1bd}
\definecolor{engi4}{HTML}{89e5b1}
\definecolor{engi5}{HTML}{78e35a}
\definecolor{engi6}{HTML}{94bb16}
\definecolor{engi7}{HTML}{b1c16a}
\definecolor{engi8}{HTML}{fed47f}
\definecolor{comp1}{HTML}{ff5319}
\definecolor{comp2}{HTML}{ff4265}
\definecolor{comp3}{HTML}{ff52a4}
\definecolor{comp4}{HTML}{eb72d8}
\definecolor{comp5}{HTML}{ba90fa}
\definecolor{comp6}{HTML}{80a8ff}
\definecolor{comp7}{HTML}{4db8ff}

\NewDocumentCommand{\statcirc}{ O{#2} m }{%
    \begin{tikzpicture}
    \fill[#2] (0,0) circle (1.0ex); 
    \fill[#1] (0,0) -- (180:1ex) arc (180:0:1ex) -- cycle; 
    \end{tikzpicture}
}
\begin{document}

\title{\vspace{-0.5cm} Gaining Insights on Student Course Selection in Higher Education with Community Detection \vspace{-0.5cm}}
%
\numberofauthors{4}
 \author{ 
 \alignauthor
 Erla Guðrún Sturludóttir\titlenote{EGS and EA contributed equally.}\\
        \affaddr{Reykjavík University}\\
        \affaddr{Menntavegi 1}\\
        \affaddr{102 Reykajvík, Iceland}\\
        \email{erlas13@ru.is}
 \alignauthor
 Eydís Arnardóttir\\
           \affaddr{Reykjavík University}\\
        \affaddr{Menntavegi 1}\\
        \affaddr{102 Reykajvík, Iceland}\\
        \email{eydis13@ru.is}
 \alignauthor Gísli Hjálmtýsson\\
          \affaddr{Reykjavík University}\\
        \affaddr{Menntavegi 1}\\
        \affaddr{102 Reykajvík, Iceland}\\
        \email{gisli@ru.is}
 \and  
 \alignauthor María Óskarsdóttir\titlenote{Corresponding author.}\\
           \affaddr{Reykjavík University}\\
        \affaddr{Menntavegi 1}\\
        \affaddr{102 Reykajvík, Iceland}\\
        \email{mariaoskars@ru.is}
        }

\maketitle


\begin{abstract}
Gaining insight into course choices holds significant value for universities, especially those who aim for flexibility in their programs and wish to adapt quickly to changing demands of the job market. However, little emphasis has been put on utilizing the large amount of educational data to understand these course choices. Here, we use network analysis of the course selection of all students who enrolled in an undergraduate program in engineering, psychology, business or computer science at a Nordic university over a five year period. With these methods, we have explored student choices to identify their distinct fields of interest. This was done by applying community detection to a network of courses, where two courses were connected if a student had taken both. We compared our community detection results to actual major specializations within the computer science department and found strong similarities. To compliment this analysis, we also used directed networks to identify the "typical" student, by looking at students’ general course choices by semester. We found that course choices diversify as programs progress, meaning that attempting to understand course choices by identifying a "typical" student gives less insight than understanding what characterizes course choice diversity. Analysis with our proposed methodology can be used to offer more tailored education, which in turn allows students to follow their interests and adapt to the ever-changing career market.
\end{abstract}

%


\section{Introduction}
University students enter higher education with a plethora of courses to choose from on their path to graduation. Gaining insight into student choices holds significant value for universities, especially those who aim for flexibility in their programs and those who wish to adapt quickly to changing demands of the job market. For example, the fast rise in popularity of machine learning over the past years could impel universities to make machine learning and related courses readily available to their students. In contrast, more subtler trends could be directly identified by the students' choices rather than an obvious shift in the job market.

Numerous studies based on questionnaires and surveys have found that there are various components that contribute to a student’s course selection \cite{babad_experimental_2003, maringe_university_2006, milliron_exploring_2008}. These are factors such as learning value, workload, age, academic performance and the professor’s lecturing style \cite{babad_experimental_2003}. Of these, the learning value of the course (which refers to factors such as intellectual level and interest in the topic) has been found to be the most influential factor in course selection. Course selection has also been a target in studies aiming to understand the gap between student mindsets and career demands \cite{milliron_exploring_2008}. Maringe \cite{maringe_university_2006} found that although intrinsic interest was important, course choices depend mainly on future career goals. According to the author, universities may need to adapt their strategies to the idea that students’ course choices now seem to reflect their expectations of future employment rather than simply interests. With that in mind, universities would benefit greatly from a deeper understanding of the path their students choose as they work towards their degree.

Educational data mining (EDM) has risen as a new field to answer these and other questions about students and their learning environment. It utilizes a variety of analytical methods and applies them to the vast amounts of data that has become available with increased digitization of administrative educational information. For example, EDM methods have already been applied to try to accurately predict college success using common classification algorithms with different feature sets \cite{yu_towards_2020}. They have also been used to analyze student clicking behavior in online courses to determine students’ learning strategies and how those strategies can have an impact on their learning outcomes \cite{akpinar_analyzing_2020}, as well as to predict student dropout \cite{deeva2017dropout}. One area of educational studies that has not received much attention is student course selection, despite its importance in understanding student interests and preparing them for a future career \cite{turnbull_entropy_2020}. 

In this paper, we aim to reveal patterns in course selection through EDM, providing a new data-driven technique based on institutional analytics to gain insight into students' interests that would otherwise be difficult to discern. We examine whether a) network analysis, with a focus on community detection, can effectively be used to identify university students’ fields of interest, and b) whether we can identify a typical student of a given major using directed networks, to contrast with the specific fields of interests. To accomplish this, we use a weighted projection network in combination with community detection to explore student course selection. We focus on communities of elective courses for different majors and compare them to some of the official specializations the university already has to offer. We compliment this with an analysis of a “typical” student's path to graduation. Deeper understanding of students' choices is a stepping stone into allowing students to take more control over their studies, improve flexibility in the curricula, and facilitate students' pursuit of their interests. 

The rest of this paper is organized as follows: In the next section, we review related work. We follow this with a detailed account of our methodology, and then present the results for each of the proposed research questions. We finish with a discussion of our discoveries and suggest future directions for this work.

\section{Related Work}
A promising method for EDM is to represent educational data as networks. In general, networks consist of nodes and edges, where the nodes can for example represent people, countries or cells \cite{fortunato_community_2010}. Edges represent the connections between the nodes and can be based on factors such as spatial and temporal proximity or social connections such as a friendship between two people, the spread of a deadly virus in two countries, or a synapse between two neurons \cite{borgatti_network_2009}. Network analysis is used to look at internal characteristics and the connections and patterns of nodes and edges, providing the ability to better understand the fundamental structure of networks and the real-life phenomena they model \cite{wellman_network_1983}.
 
A common application of network analysis in educational settings is to understand social connections between students. This has helped reveal the negative effects of student interdependence in music education programs and its relationship to the program's friendship networks \cite{sarazin_can_2017}, as well as identifying how positive and negative friendship ties emerge \cite{sarazin_disliking_2021}. Network analysis has also helped clarify the relationship between students' social networks and the development of their academic success \cite{blansky_spread_2013, gitinabard2018your}. Furthermore, looking at students' social networks over time, close coequal communities are typically formed early on \cite{xu2018many}, although in some cases, students enhance their performance due to social relations outside their assigned group \cite{rienties2018turning}.

Different methods can be used to analyze networks, for example by looking at structural characteristics such as centrality, which indicates the importance of any given node in the network by assuming nodes that are more central have higher control over information passed through the network \cite{borgatti_network_2009}. Another common way of analyzing networks is through community detection. Community detection is a method that allows for the aggregation of different nodes into communities based on shared characteristics by identifying groups of nodes that have a high number of edges within themselves but fewer edges to other groups \cite{fortunato_community_2010}. 

There are various methods to achieve community detection, but in the current study we use the Louvain algorithm. This is an established, fast converging method that produces accurate communities with high network modularity \cite{blondel_fast_2008, que_scalable_2015}. This method performs well compared to other community detection algorithms, especially on smaller networks such as the one we analyze here \cite{lancichinetti_community_2009}. It successfully balances accuracy and computational complexity. The Louvain algorithm has been applied in various fields, for example to help identify communities of intrinsic brain systems that show consistent functional connectivity both during rest and task performance \cite{cole_intrinsic_2014}, or to help create friend lists for Facebook users \cite{liu_analyzing_2011}. 

Although students' social networks have been studied, the exploration of students' course choices through network analysis has only a few precedents. Within the EDM field, Kardan et al. \cite{kardan_prediction_2013} used neural networks to predict course enrollment based on various factors such as course and instructor characteristics, and course difficulty. Further, Turnbull and O'Neale \cite{turnbull_entropy_2020} used network analysis with community detection and entropy measures to explore course enrollment in STEM subjects at the high school level. Among other results, they revealed that indigenous populations showed higher levels of entropy in their enrollment patterns, which was moderated by adolescent socioeconomic status. Neither of these studies focused on detecting fields of interests from course selection patterns.

\section{Methods}
\subsection{Data Source}
Here, we use student and course data from a Nordic university (NU). The university offers many different areas of study, including preliminary studies, undergraduate and graduate degrees. Most NU students are undergraduate students, and the NU undergraduate programs also offer the most variety of courses. Generally, the majority of NU undergraduate programs' courses are mandatory. These are the core courses each department decides is essential to their study program. The rest of the courses are either free choice electives, which can be any course in the university that the student qualifies for, or restricted elective courses from a selection tailored to the specific major. 

We sample data from all graduated NU students that enrolled in the year 2014 or later and completed undergraduate programs in engineering, psychology, business, or computer science (CS) before 2021 (the total number of students was 1481). The university offers other programs as well, but we left them out since they have fewer students. The variables we look at include the student's registration ID and registration semester, the name and semester of each course a student has completed, and whether they passed or failed the course. We also include each student's department, major, and type of study (undergraduate, graduate, etc.)

\subsection{Data Cleaning}
To anonymize the data, we remove anything that could identify the students, specifically their social security number and a numerical registration ID. We give each student a unique random sequence of numbers to replace both original numbers. For each student, we also remove any courses that they had registered for and then de-registered early within the same semester. Further, for each major, courses taken by fewer than 5\% of students are considered outliers and removed. 

\subsection{Network Analysis}
\subsubsection{Bipartite networks}
We apply network analysis to the data to explore the fields of interests of NU students from a data driven perspective. Many real-world networks have a bipartite structure, where nodes belong to one of two groups or divisions. Any edge in the network must connect nodes of opposite groups and there are no within-group edges \cite{banerjee_properties_2017}. In our bipartite network, the students make up one division of the nodes, and courses the other division. If a student has taken a course, an edge is created between the respective nodes. Since edges represent that a student has taken a course, there will be no edge between two students and no edge between two courses (see Figure 1, left). 

Although bipartite networks give a more realistic and detailed representation of the system, analyzing them can be complex. Therefore we project the bipartite network onto its unipartite counterpart (see Figure 1, right) \cite{banerjee_properties_2017}. This leaves a network with one type of nodes that can be analyzed with typical network methods. The resulting projected network consists of nodes representing the courses and edges between two nodes indicating that a student has taken both courses.

A base problem with projection of bipartite networks is that a lot of important information in the original bipartite network is lost. Thus, we may end up connecting all courses in the network to each other --and form a clique-- as long as they have at some point been taken by the same student, without taking into account how many students connected the two courses in the original bipartite network. Here, we address this by assigning weights to the edges in the projected network \cite{banerjee_properties_2017}, where the weights represent the number of students who have taken both courses (see Figure 1). 

\begin{figure}
    \includegraphics[width=0.47\textwidth]{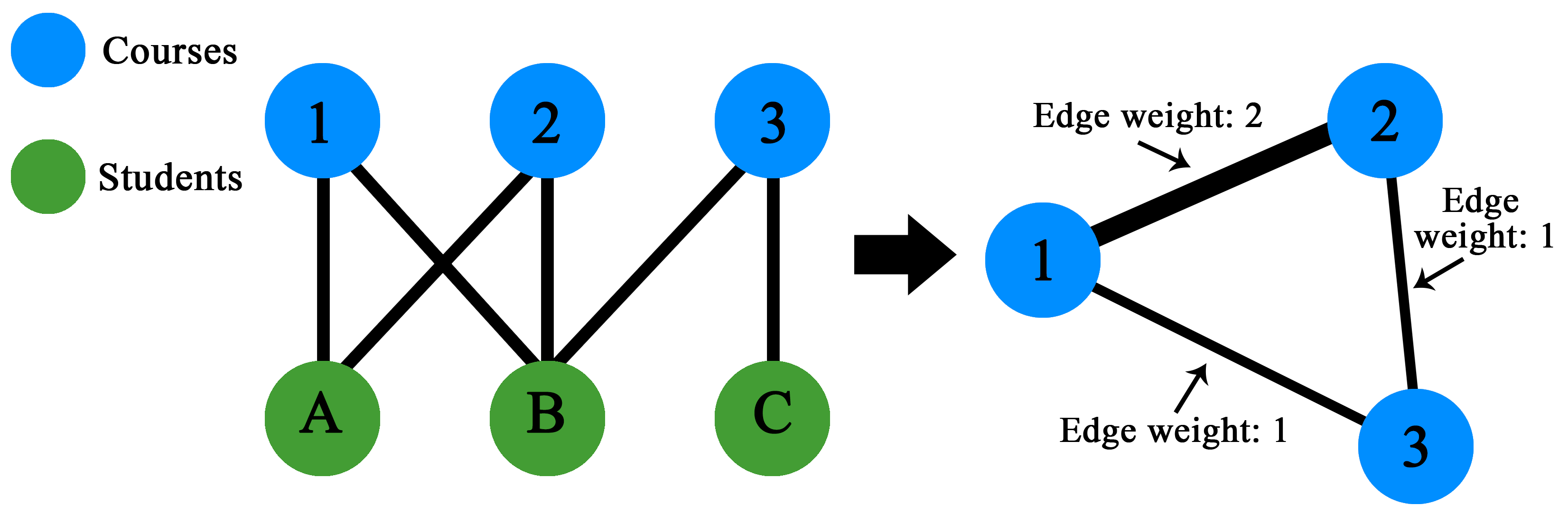}
    \centering
     \caption{From bipartite network to weighted projected network. Left: a bipartite network, where the blue nodes represent courses and the green nodes, students. Right: a unipartite network has been obtained from the bipartite network, where the nodes are courses and the edges have weights that determine how many students have taken both courses. }
\end{figure}

\subsubsection{Community detection}
Building on the weighted projected networks, we use community detection with the objective of inferring fields of interests in students' course selection. To identify fields of interest, we want to emphasise electives. However, in our data set, the information on which courses are mandatory and which are electives is incomplete. Mandatory courses along with very popular electives appear in the network as hubs, which usually occur in real-world networks as nodes with much higher degrees and edge weights than the other nodes \cite{barabasi_network_2013}. 
We define hubs in two ways
\vspace{-0.5cm}
\begin{enumerate}
\itemsep0em
    \item Data-driven (DD): a node is a hub if its total edge weight is at least one standard deviation above the edge weight mean of all nodes.
    \item Ground truth (GT): a node is a hub if it is a mandatory course\footnote{In our data set, the information on which courses were mandatory and which were electives was incomplete, and had to be filled in manually.}.
\end{enumerate}
\vspace{-0.5cm}
We use both definitions to remove hubs and therefore obtain two distinct networks.
 
Having removed hubs, we apply the Louvain algorithm for community detection on the resulting distinct networks \cite{blondel_fast_2008}. This method is computationally efficient and is based on optimizing partition modularity, which is a measure of edge density within a partition (or proposed community) as opposed to edge density between partitions \cite{fortunato_community_2010}. Higher modularity therefore suggests a more cohesive community, separate from the others in the network. Importantly for our analysis using weighted projected networks, the Louvain algorithm can be used both with weighted and unweighted edges. As demonstrated in Figure 2, the method starts by assigning each node to its own community \cite{blondel_fast_2008}. It then iterates over all nodes of the network and assesses the modularity gain obtained by assigning the node to the same community as each of its neighboring nodes. Next, the node is assigned to the community that yields the largest positive modularity gain, or maintains its current community if no positive modularity gain can be achieved by switching communities. This way, each new community assignment brings us closer to optimal modularity. The nodes are usually considered multiple times and the final iteration is determined when no switch leads to a gain of modularity, resulting in optimal partitioning of the network. This optimal partitioning is a local maxima, as the result is influenced by which node is considered first and the order in which nodes are visited. For some communities, we re-apply the Louvain algorithm for more detailed results, while using the inter/intra weight density ratio described below to ensure our communities maintain high quality.

\begin{figure}[h!]
    \includegraphics[width=0.47\textwidth]{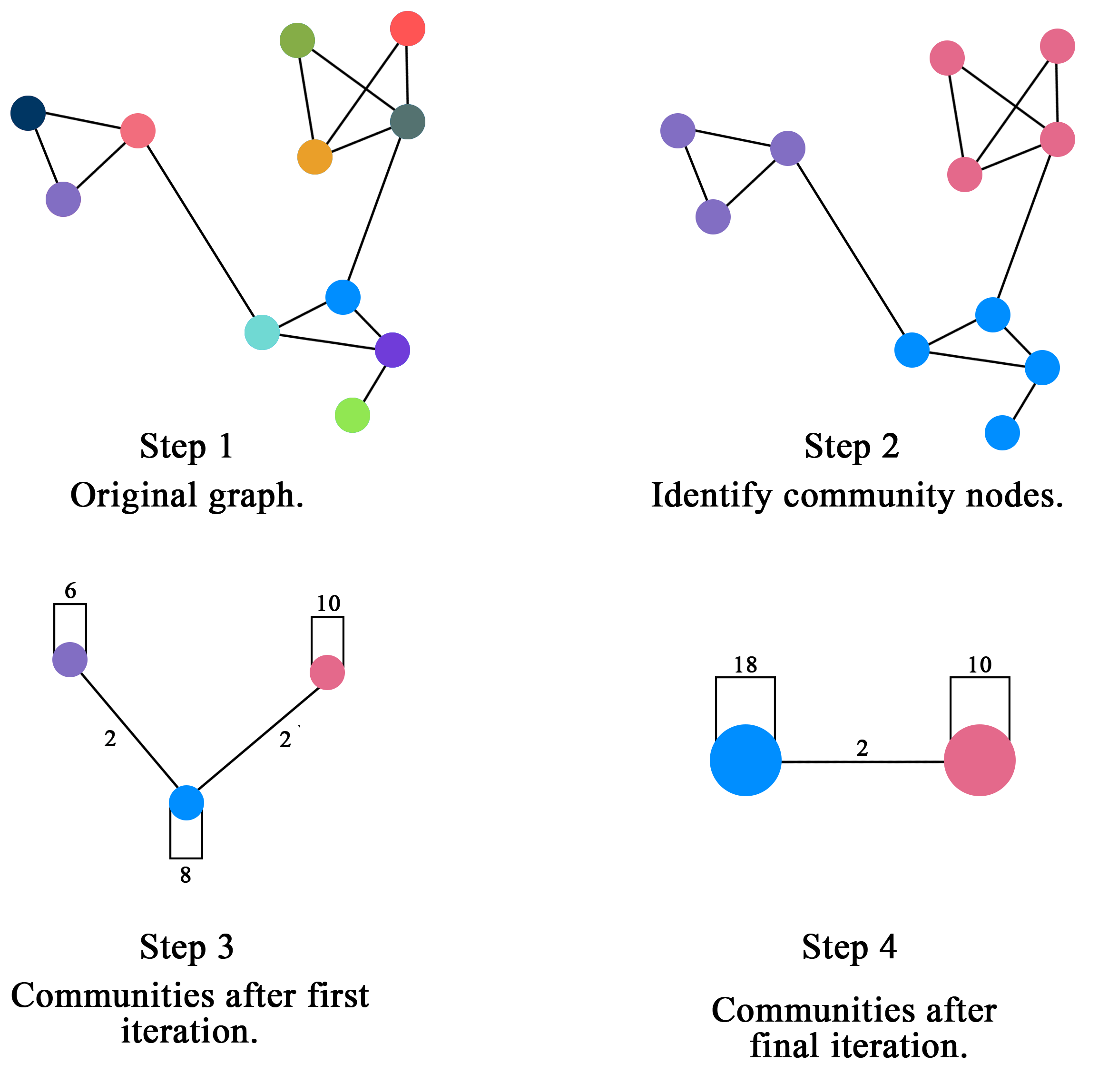}
     \caption{The Louvain algorithm. The first step of the algorithm is to assign each node to its own community. In step 2, a random node is selected to start the community aggregation process. All nodes are visited and allocated to the community of one of their neighbors or maintain their current community, depending on which choice gives the highest gain in modularity for the network. When no more modularity gain is possible in the network, step 3 is to aggregate the nodes of each community into new super-nodes. Here, the numbers given show the sum of node edges within and between supernodes. Steps 2 and 3 are then repeated until modularity has been optimized, as seen in step 4. }
\end{figure}

\subsubsection{Community validation}\label{subsec:CommunityValidation}
Although the objective of community detection is to split nodes into groups based on their connections within versus outside the group, there are many more aspects to consider \cite{fortunato_community_2010}. One important factor is intra-cluster density, which refers to how many edges there are within the community as a ratio of how many possible edges there could be if all nodes of the community were connected to each other. This is contrasted by inter-cluster density, which shows how many edges go from the community to the rest of the network as a proportion of the maximum possible connections. High intra-cluster density may suggest a strong and cohesive community, however if it coincides with equally high inter-cluster density, it may simply suggest a strong and cohesive overall network.

To assess the quality of our communities, we use intra and inter weight density \cite{fortunato_community_2016}. This is the same as intra and inter edge density previously described, but now accounting for weighted edges. The two are defined as follows:
\begin{equation}
\mathit{WD}_{\text{inter}} = \frac{w_C^{\text{ext}}}{\bar{w}n_C(n-n_C)}
\label{eq:inter_weight_density}
\end{equation}
     and 
\begin{equation}
\mathit{WD}_{\text{intra}} = \frac{w_C^{\text{int}}}{\bar{w}n_C(n_C - 1)/2}.
\label{eq:intra_weight_density}
\end{equation}

In Equation \ref{eq:inter_weight_density},
$w_C^\text{ext}$ is the sum of edge weights connecting the community to the rest of the network, or external community edges. We divide this by the estimated total edge weight of the network, which shows the edge weight of edges going from the community to the rest of the network as a proportion of the maximum possible edge weight (assuming that the average edge weight of the fully connected network were unchanged). Here, $\bar{w}$ is the average edge weight of the network, $n$ is the total number of nodes in the network and $n_C$ is the total number of nodes within the community. Similarly, in Equation \ref{eq:intra_weight_density}, \text{$w_C^\text{int}$} refers to the sum of edge weights inside the community, which is divided by the expected total edge weight within the community. We then use a ratio of these two measures ($WD_{\text{inter}}$ / $WD_{\text{intra}}$) to obtain the community strength on a scale where 0 is the strongest possible value, meaning that the community would be disconnected from the rest of the network, and a value of 1 would mean that the community is equally connected within itself as to the rest of the network. We use this ratio not only to determine the community strength, but also to ensure that as we create smaller and more focused communities, community strength is not compromised. 

We use clustering similarity to compare communities discovered in the networks with hubs removed in the two distinct ways. Clustering similarity was first proposed by Gavrilov et al. \cite{gavrilov_mining_2000} and has been used by others for cluster validation \cite{lismont_closing_2019, lin_rotation-invariant_2012}. Importantly, it does not assume that the number of communities is equal for both methods. This is essential as the community detection method we use does not produce a predefined number of communities. The similarity between two communities is calculated by the number of courses common to both communities relative to the total number of courses in both communities. With this we compare networks where hubs were removed using the DD method and networks where hubs were removed using the GT method. Clustering similarity is given by
\begin{multline}
\text{Similarity} (G,C) = \frac{1}{k} \sum^k_{i=1}\max_{j \in [0,l]} \text{Sim}(G_i, C_j) \\
\text{with} \quad \text{Sim}(G_i,C_j) = \frac{2\left|G_i \cap C_j \right|}{\left| G_i\right| + \left|C_j\right|},
\end{multline}
where \textit{G} are all communities detected after using the GT method while \textit{C} are all communities detected after using the DD method and \textit{k} and \textit{l} are the numbers of communities from the GT and DD methods, respectively. For each GT community, we iterate through all communities found in networks filtered by the DD method and look for the maximum similarity. Clustering similarity ranges from 0 to 1, with 1 representing identical communities.

\subsubsection{Comparing communities and real NU specializations}
To further assess the real-world application of the communities we detect, we compare them to specializations within NU’s Computer Science (CS) department. Any student who pursues an undergraduate degree in CS at NU has the option to graduate with a specialization in a certain field. The specializations do not need to be declared at enrollment but any student who fulfills the requirements can choose to add this to their graduation certificate. The specializations offered are Artificial Intelligence, Law, Web- and User Experience (UX) Design, Sports Science, Psychology, Game Development and FinTech. Each specialization has 2-4 core courses that students need to complete, along with 1-3 courses from a pool of specialization-specific electives. Our approach to defining fields of interest is purely through data driven community detection. Comparing the detected communities with these specializations helps validate the results and perhaps provide a reference for the creation of new specializations. 

We compare not only the courses in each community and specialization, but also the number of students belonging to a specialization versus those belonging to the corresponding community. We define a student as belonging to a community if they have taken at least 50\% of the community's courses, with a special case of two course communities where both courses have to be completed.

\subsubsection{Directed network}
In our analysis above, we have only used an undirected course network as our community detection is indifferent to the order in which the students take these courses. However, one of our objectives is to use network analysis to define a typical student, and how he/she progresses through their studies. We do this by creating a new network, where each node is characterized by the set of courses a particular student has taken in a particular semester (see Figure 3). For example, all students who took the exact same courses in their first semester share a “first semester” node. We determine the weight of the node by the number of students who took these courses during their first semester. This is repeated for each semester. Then, for each student we add a directed edge between nodes representing their courses in consecutive semesters. Thus the nodes tell us which courses a student took in a particular semester and the direction of the edges tells us the order of the semesters. The weight of edges is determined by the number of students who took the semesters in that order. The nodes with the highest weight for each semester provide an idea of how the “typical” student organizes their studies, while the community detection previously described illustrates how real students can differ from the average in predictable ways.

\begin{figure}[htp]
    \includegraphics[width=0.47\textwidth]{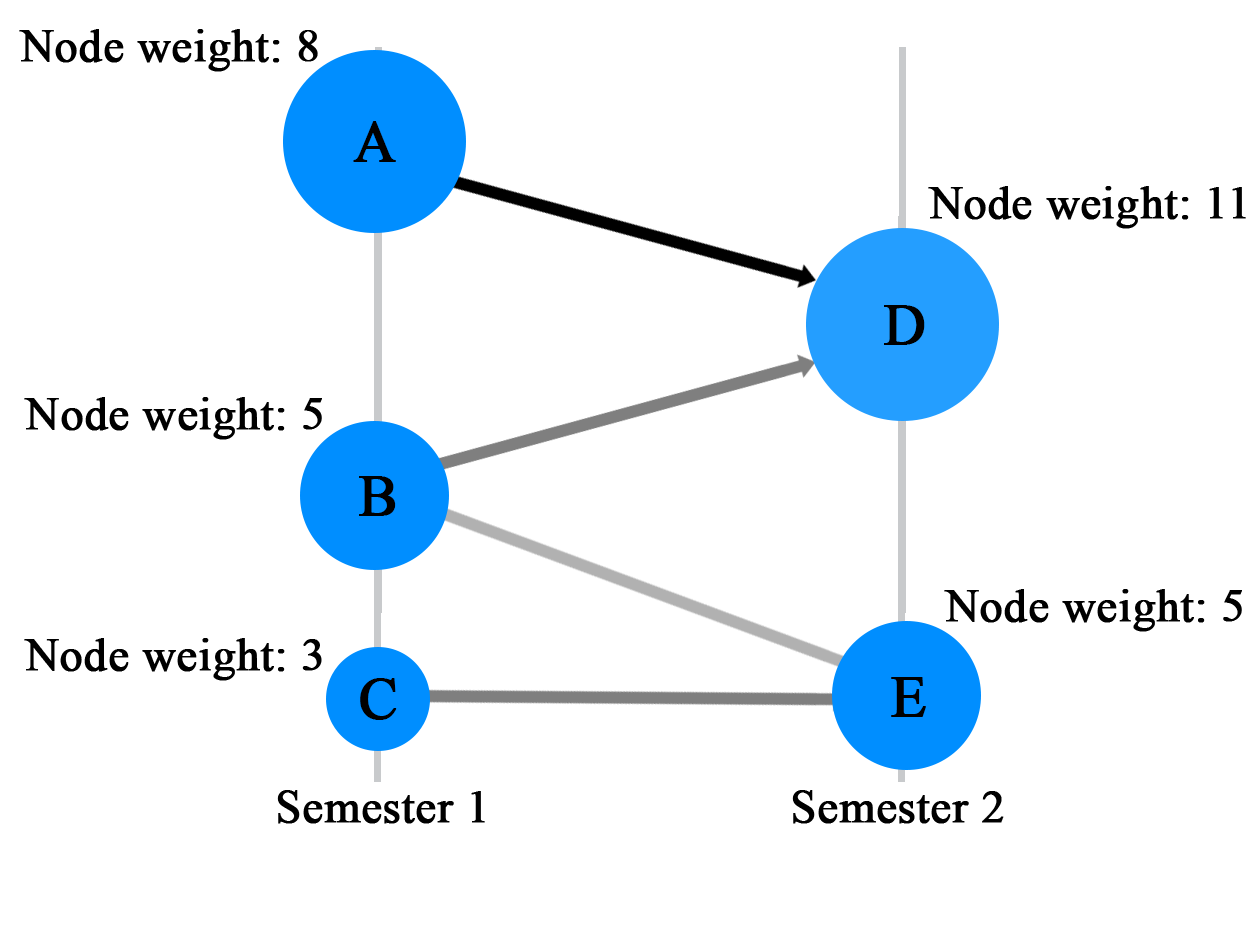}
     \caption{An example of a semester network setup over two semesters. A node in the network is made up of a collection of courses one or more students took during a semester, while the weight of the node is determined by the number of students who have taken that semester.}
\end{figure}

\subsection{Tools}
Aside from the initial retrieval and anonymization of data, which we do using C\# and SQL, all code for the data analysis was written in Python 3.9. We use multiple Python libraries to help with the data analysis. To perform our network analysis, we mainly utilize the NetworkX library \cite{team2014networkx}. For more general data manipulation, we use the pandas library \cite{reback2020pandas}. We used the Gephi software package for the majority of our network visualization \cite {ICWSM09154}, along with the Matplotlib library \cite{Hunter:2007}.

\section{Results}
\subsection{Strong Communities that Reflect Real-Life Fields of Interest}

We conducted community detection with the Louvain algorithm on four undergraduate majors: engineering, psychology, business, and computer science (see Figure 4). These majors have quite different program structures and emphases on electives, with the business major having the lowest number of elective courses allowed in their study plan (four electives). This is followed by the psychology major with five electives, then CS (11 electives) and finally engineering, which offers only four free electives but nine "guided electives" (that is, nine electives must be specific to engineering), depending on the chosen engineering specialization. We compare the effectiveness of removing hubs versus removing mandatory courses on the community detection results for each major using clustering similarity. For these majors, we observe strong clustering similarity ranging from 0.66-0.95, with psychology having a clustering similarity of 0.66 and CS a clustering similarity of 0.95. This indicates that hub removal can substitute the removal of actual mandatory courses effectively. 

We first look at the communities for the engineering department, which after hub removal consisted of 77 courses taken by 496 undergraduate students. The Nordic university offers various undergraduate engineering programs such as biomedical engineering, financial engineering, and mechatronics engineering. These engineering majors all fulfill the same core courses in addition to some additional major-specific requirements. These majors are quite structured and offer few free elective courses. Due to the similarity in the core courses of these programs, we group them together into a more general engineering major. This means that the hub removal method removed general core engineering courses but leave most specialty-specific courses in the network. The resulting engineering network has 77 course nodes and 2614 edges. The clustering similarity is 0.75 and the weighted average inter/intra weight density ratio is 0.24. This suggests that hub removal was effective and the average community is relatively strong. The communities we have detected were eight in total as seen in Table 1. Note that communities are named after common characteristics between the majority of the courses, even though rarely all courses of a community fall within that definition. As expected, these communities mainly correspond to the official engineering majors such as financial, biomedical, and electrical engineering, with electrical engineering being our strongest community ($WD_{\text{inter}}$ / $WD_{\text{intra}}$ = 0.05). However, we also observe unrelated communities that supersede the official majors, such as a community of applied design and another for business related courses not mandatory in the financial engineering major. Courses in these communities are commonly taken together by engineering undergraduates, suggesting a common interest not credited to the specialized majors. 

The psychology major, which consists of 274 students, is one of the more structured majors we looked at, offering only five electives. The network is comprised of 30 nodes and 346 edges. We detected five psychology communities with a clustering similarity of 0.66 and a weighted average $WD_{\text{inter}}$ / $WD_{\text{intra}}$ = 0.25, again demonstrating strong communities but slightly less effective hub removal than for the engineering major. Because psychology offers fewer electives, and these electives are quite diverse, the community themes observed are more vague than those of the engineering majors. For example, the strongest community identified was a community of popular courses (Popular Courses I in Table 2). This community has an $WD_{\text{inter}}$ / $WD_{\text{intra}}$ = 0.2 and consists of common electives and a few core courses not detected during hub removal. Despite the community themes being less clear than for the other majors, we do find an interest in the sub-fields of sleep, child and behavioral psychology, and management. With a higher variety in available psychology electives, it would be interesting to see whether stronger interest fields emerge.

There are 334 undergraduate students in our data set who majored in business. For this major, the network consists of 36 course nodes and 504 edges, with a clustering similarity of 0.84. The weighted average $WD_{\text{inter}}$ / $WD_{\text{intra}}$ is 0.25, again suggesting strong communities. This is not unexpected, as the business major only allows electives in the final year, giving business students less room to pursue distinct interests outside their core subjects. Table 3 shows the five communities identified within the business major. As with psychology, the strongest community is that of popular courses, which includes the most common electives in the business majors along with a handful of newer core courses ($WD_{\text{inter}}$ / $WD_{\text{intra}}$ = 0.07). These core courses were recently added to the study plan, meaning that they were only mandatory for a minority of the students in our data set. This is why these core courses were not identified as hubs and removed during hub removal. 
The business major also contains the weakest community of all the majors, management ($WD_{\text{inter}}$ / $WD_{\text{intra}}$ = 0.71). As the  name suggests, this community includes various courses on management, such as service management and project management. The low inter and intra weight density ratio is interesting, as intuitively these courses would seem very connected. This is why measuring community strength is vital in determining the importance of the detected communities. The other business communities are both strong and reflect more specific interests, suggesting that there are students of the business major who actively seek distinct interests despite the program having no official specializations. 

The last major we explore is CS, with 377 students. Computer science has the least structured study plan of the four majors, as it puts a higher emphasis on unstructured flexibility and free electives. The CS course network consists of 59 nodes and 1492 edges. The communities are the strongest we found, with a weighted average $WD_{\text{inter}}$ / $WD_{\text{intra}}$ of 0.21 and a clustering similarity of 0.95. Most, but not all, detected communities seem to reflect an interest in a CS sub-field. However, the strongest community we have discovered was Deprecated Courses I (see Table 4), which represents older courses that may have been core courses at some point but are no longer being offered ($WD_{\text{inter}}$ / $WD_{\text{intra}}$ = 0.08). We conjecture that this community exists as some older students re-register to complete their undergraduate degree, for example after previously completing a CS diploma or taking a longer study break. It is therefore very intuitive that this specific sub-field is combined into our strongest community. Aside from communities based on deprecated courses, the other communities suggest that there is in fact an underlying pattern of interest fields present in the CS major, as observed for the other majors explored here.

\subsection{Comparing Communities to NU Computer Science Specializations}

As a final validation of the communities we have detected for the CS undergraduate major, we now cross-reference our results with the actual specializations available for CS students. Unlike the other majors, CS offers a number of specializations meant to aid students in pursuing a specific sub-field (see Box 1 for a short description of each specialization). However, only a subsection of students choose to do this. Of the students who graduated between 2014 and 2020, inclusive, only 9.5\% fulfilled the requirements for a specialization. A further 13\% partially fulfilled a specialization's requirements, by completing at least 60\% of the specialization's core courses and 60\% of the restricted electives needed. 

Comparing the specializations and the communities we detected (shown in Table 4), we find interesting similarities. Our community detection reveals that some communities are consistent with the specializations, but there is no absolute match. For the AI specialization (taken by 11 students, or 29\% of those who graduated with a specialization), there is a partially corresponding community that includes both of the AI core courses (Artificial Intelligence and Machine Learning). There are 28 students who belong to this community, making it more popular than the official AI specialization. Although this community does not include any of the other courses from the specialization, it does include more theoretical and academically demanding courses than most other communities, suggesting a reflection of interest in theoretical computer science in general rather than specifically AI.

To fulfill the official AI specialization requirement, students must complete two core courses and three or more courses from a list of specialization-specific electives. However, in our data set most of these other electives were removed during either data cleaning (where we removed courses taken by fewer than 5\% of students) or during hub removal and are therefore not part of any community. Interestingly, two of the remaining electives overlap between the AI specialization and that of Game Development. Both these courses have been sorted by our algorithm into a community that reflects Game Development much more strongly than AI, with 67 students. This is intriguing, as we know that students are much more likely to specialize in Artificial Intelligence than 
\clearpage
   
   \begin{table}[h!]
    \caption{Community detection results for BSc in engineering.
    Clustering similarity: 0.75}
    \scalebox{1.1}{
    \begin{tabular}{|cl|c|c|}
        \hline
        & Community & No. courses & $WD_{\text{inter}}$/$WD_{\text{intra}}$ \\\hline
        \statcirc{engi1} & Comp Sci and Mechatronics &  $25$ & $0.37$ \\\hline
        \statcirc{engi2} & Engineering Management &  $15$ & $0.16$ \\\hline
        \statcirc{engi3} & Finances and Management & $10$ & $0.25$ \\\hline
        \statcirc{engi4} & Biomedical Engineering &  $10$ & $0.10$ \\\hline
        \statcirc{engi5} & Financial Engineering &  $9$ & $0.21$ \\\hline
        \statcirc{engi6} & Electrical Engineering &  $5$ & $0.05$ \\\hline
        \statcirc{engi7} & Applied Design &  $4$ & $0.29$ \\\hline
        \statcirc{engi8} & Business &  $3$ & $0.32$ \\\hline
          & \textbf{Weighted average} &   & \textbf{0.24} \\

        \hline
    \end{tabular}}
    \end{table}\vfill
        \begin{table}[h!]
        \caption{Community detection results for BSc in psychology. Clustering similarity: 0.66}
        \scalebox{1.1}{
        \begin{tabular}{|cl|c|c|}
            \hline
            & Community & No. courses & $WD_{\text{inter}}$/$WD_{\text{intra}}$ \\\hline
            \statcirc{psychpopi} & Popular Courses I &  $14$ & $0.20$ \\\hline
            \statcirc{bizpsych} & Management and Psychology &  $5$ & $0.26$ \\\hline
            \statcirc{psychpopii} & Popular Courses II &  $4$ & $0.46$ \\\hline
            \statcirc{sleep} & Sleep &  $4$ & $0.24$ \\\hline
            \statcirc{newpsych} & Developmental Psychology &  $3$ & $0.23$ \\
            \hline
              & \textbf{Weighted average} &   & \textbf{0.25} \\\hline

        \end{tabular}}
    \end{table}\vfill
    
    \begin{table}[h!]
        \caption{Community detection results for BSc in business. Clustering similarity: 0.84}
        \scalebox{1.1}{
        \begin{tabular}{|cl|c|c|}
            \hline
            & Community & No. courses & $WD_{\text{inter}}$/$WD_{\text{intra}}$ \\\hline
            \statcirc{businessblue} & Popular Courses &  $15$ & $0.07$ \\\hline
            \statcirc{businessblue2} & Management &  $6$ & $0.71$ \\\hline
            \statcirc{businessblue3} & Finance &  $6$ & $0.29$ \\\hline
            \statcirc{businessblue4} & Operations &  $5$ & $0.10$ \\\hline
            \statcirc{businessblue5} & Asset Management &  $4$ & $0.36$ \\
            \hline
              & \textbf{Weighted average} &   & \textbf{0.25} \\\hline

        \end{tabular}}
    \end{table}\vfill
 
    \begin{table}[h!] 
    \caption{Community detection results for BSc in computer science. Clustering similarity: 0.95}
    \scalebox{1.1}{
    \begin{tabular}{|cl|c|c|}
        \hline
        & Community & No. courses & $WD_{\text{inter}}$/$WD_{\text{intra}}$ \\\hline
        \statcirc{comp1} & UX and Business  &  $15$ &  $0.25$ \\\hline
        \statcirc{comp2} & Engineering &  $13$ & $0.17$ \\\hline
        \statcirc{comp3} & Web and Software &  $10$ & $0.20$ \\\hline
        \statcirc{comp4} & Artificial Intelligence &  $7$ & $0.39$ \\\hline
        \statcirc{comp5} & Deprecated Courses I &  $6$ & $0.08$ \\\hline
        \statcirc{comp6} & Game Development &  $4$ & $0.10$ \\\hline
        \statcirc{comp7} & Deprecated Courses II & $4$ & $0.23$ \\
        \hline
          & \textbf{Weighted average} &   & \textbf{0.21} \\\hline

    \end{tabular}}
    \end{table}\vfill
\begin{figure}[h!]
    \begin{tabular}{c}
    \begin{subfigure}[h!]{0.48\textwidth}
    \end{subfigure}
    \\\begin{subfigure}[h!]{0.48\textwidth}
        \centering
        \includegraphics[width=0.6\textwidth]{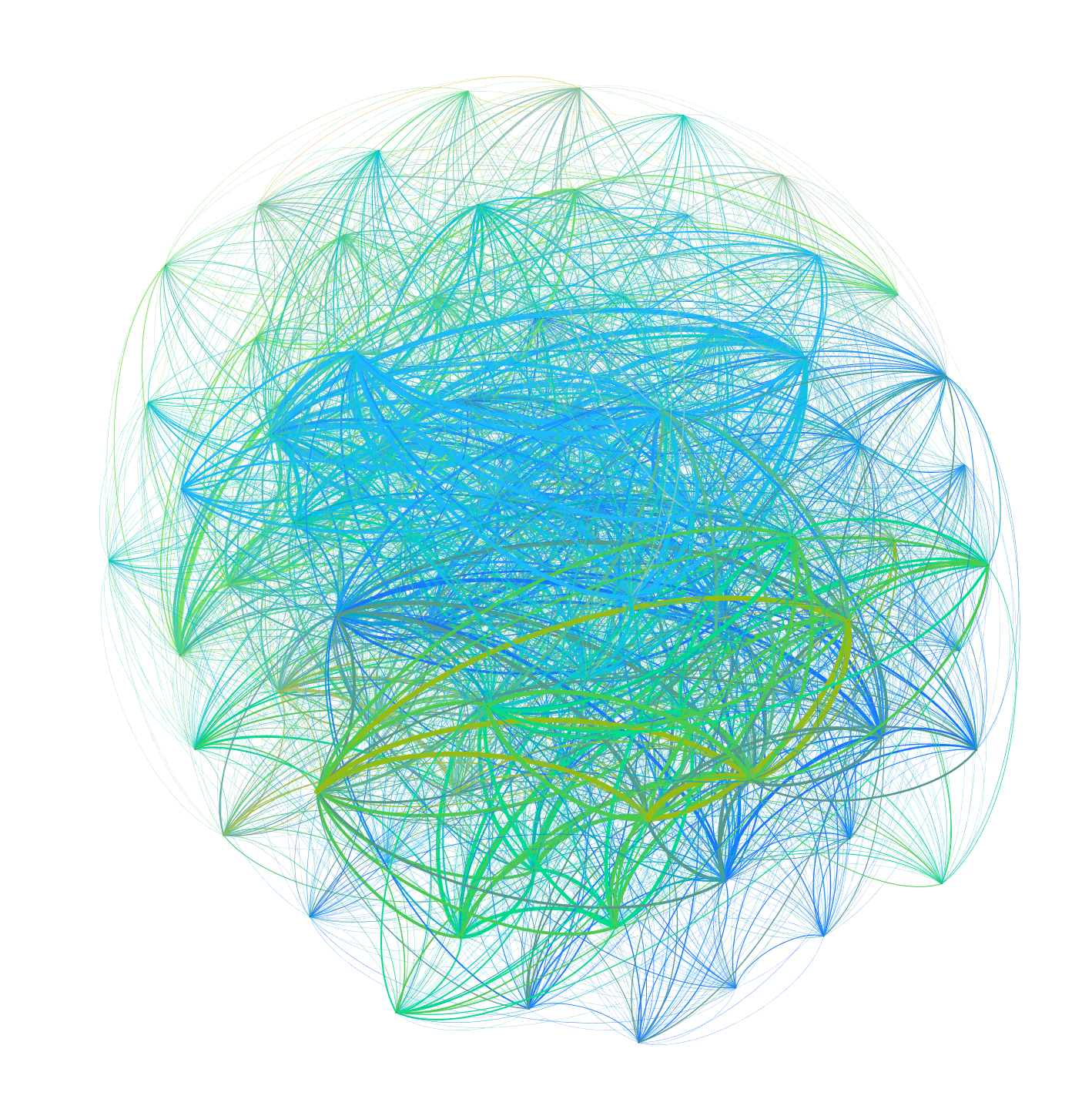}
        \caption{BSc in Engineering}
        \label{fig:bisniss}
    \end{subfigure} \\
    \\\begin{subfigure}[h!]{0.48\textwidth}
        \centering
        \includegraphics[width=0.6\textwidth]{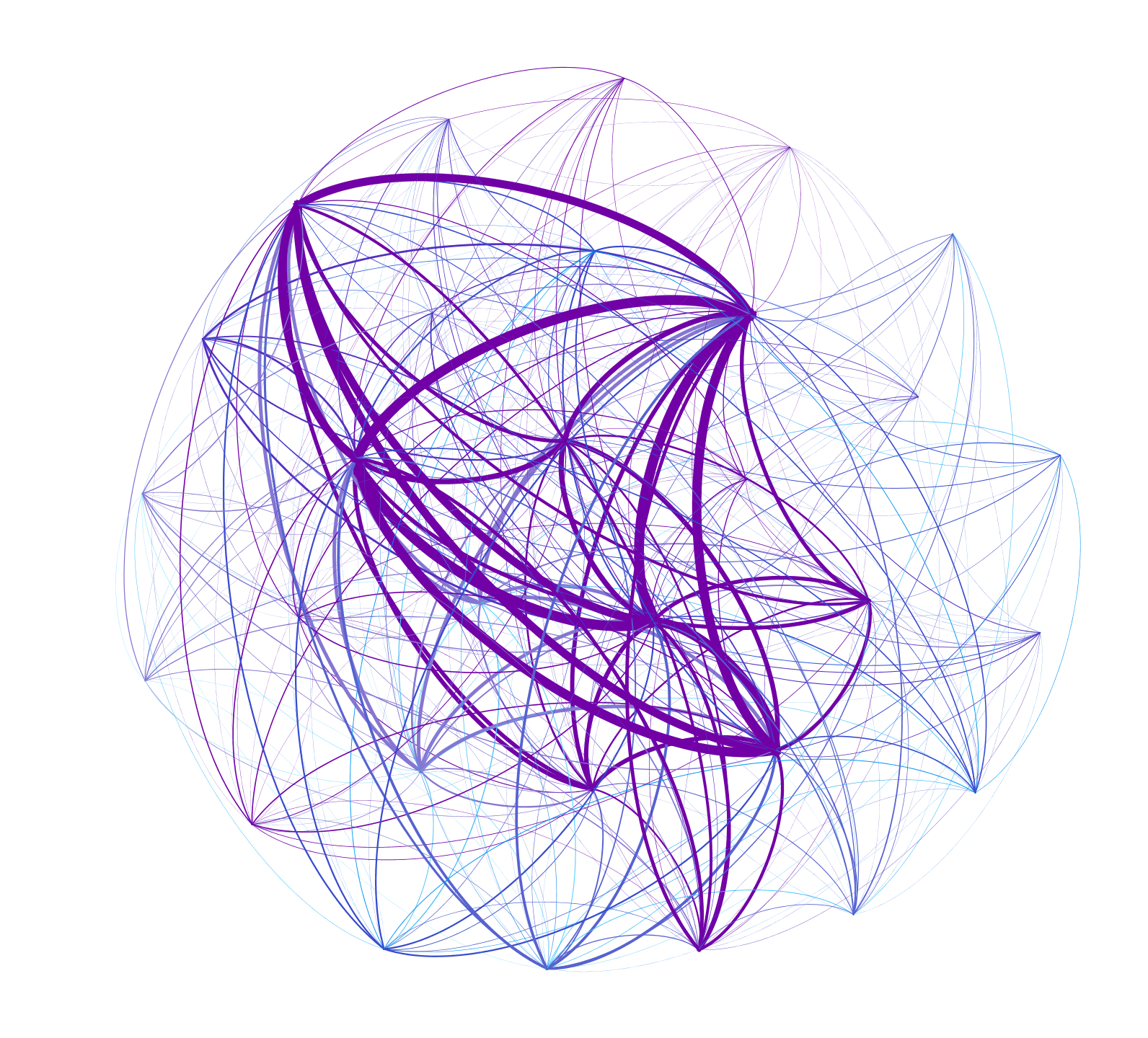}
        \caption{BSc in Psychology}
        \label{fig:bisniss}
    \end{subfigure} \\
    
    \\\begin{subfigure}[!h]{0.48\textwidth}
        \centering
        \includegraphics[width=0.8\textwidth]{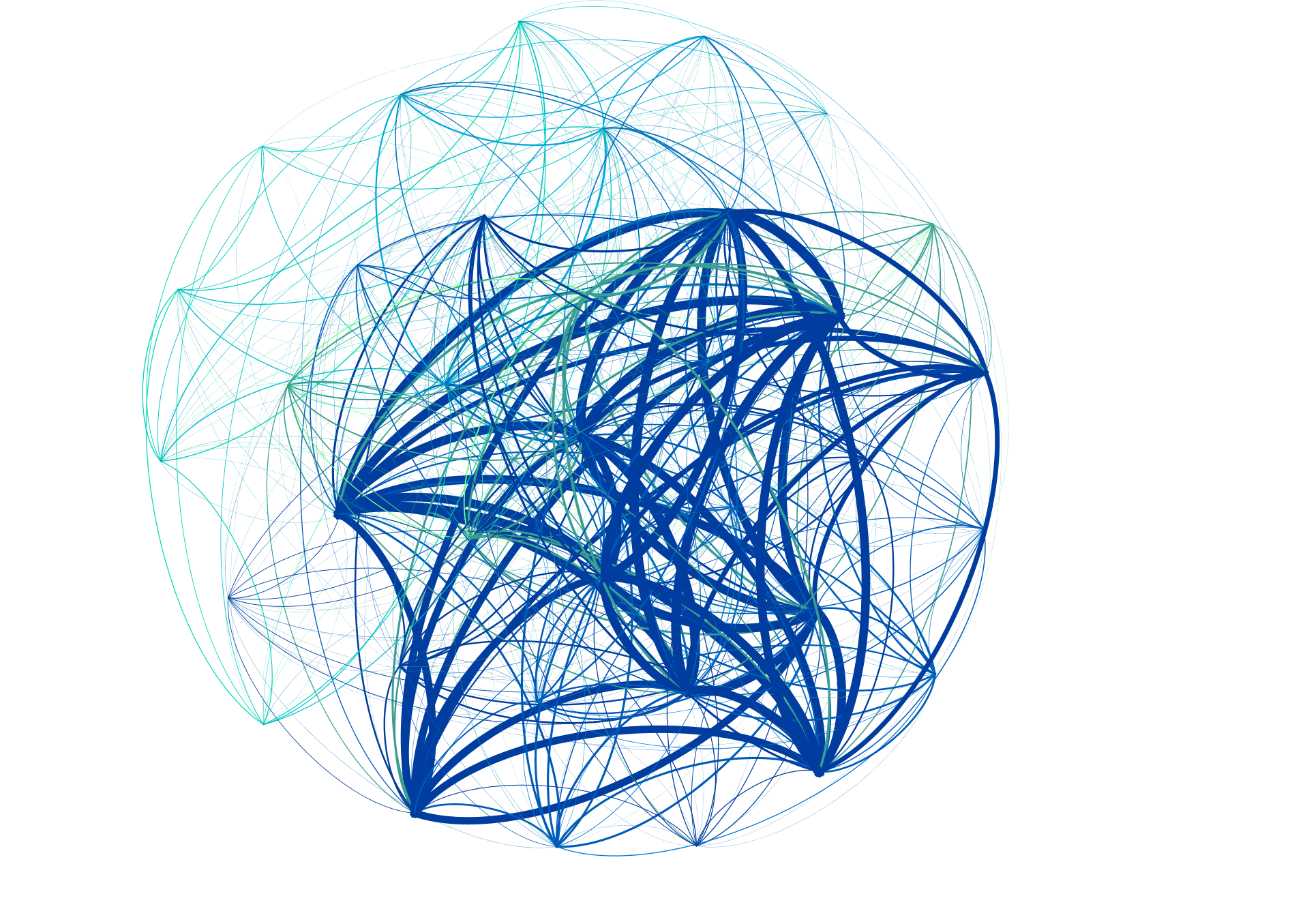}
        \caption{BSc in Business}
        \label{fig:bisniss}
    \end{subfigure} \\

    \begin{subfigure}[h!]{0.48\textwidth}
        \centering
        \includegraphics[width=0.6\textwidth]{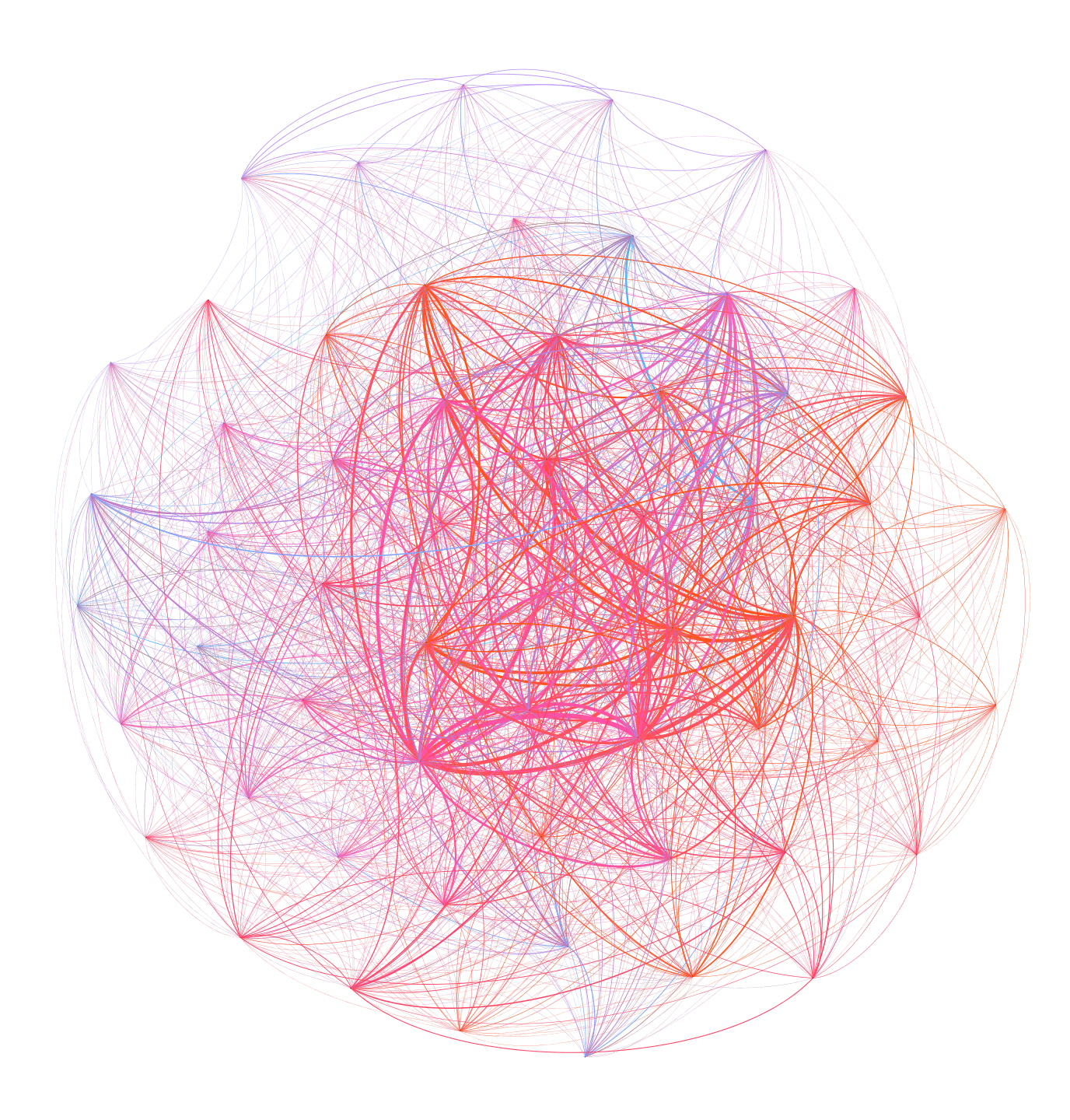}
        \caption{BSc in Computer Science}
        \label{fig:bisniss}
    \end{subfigure} \\
\end{tabular}
\caption{A network for each major after community detection, these networks correspond to the tables on the left}
\label{fig:big_picture}
\end{figure}
\clearpage
Game Development (only one student in our data set fulfills the requirements for Game Development), but this indicates that the gaming sub-field of Artificial Intelligence may be the biggest area of interest for these students.

The final specialization for which we discovered a similar community is Web and UX design, which was by far the most popular specialization taken by students (with 23 students, or 64\% of all students who had a specialization). While this specialization encompasses both web programming and user experience, the corresponding community of Web and Software Development (with 84 students) is much more web than UX specific. Most of the UX related courses belong to a separate community of 21 students that unites UX and business rather than UX and web design. This suggests that dividing the Web and UX design specialization into two distinct specializations (Web design and UX design) might be more appealing to students.

\noindent\fbox{%
    \parbox{\linewidth}{%
    \textbf{BOX 1. OFFICIAL CS SPECIALIZATIONS}
        \vspace{\baselineskip}  

        \textbf{Artificial intelligence:} Core courses reflecting an interest in AI and machine learning, with electives focused on game development and analytical skills.
        
        
        \textbf{Game design:} Core courses encompass game development in general, computer graphics and game engine architecture. Electives reflect more general programming skills and AI.
        
        
        \textbf{FinTech:} Both core courses and electives focus on the financial part of the Financial Technology discipline, as all students taking these courses gain software development skills from the core courses of the CS major. 
        
        
        \textbf{Web and UX design:} As the name suggests, most courses for this specialization directly relate to either web programming (such as the courses Web Programming II and Web Services) or user experience (User-Focused Software Development, Human-Computer Interaction).
        
        
        \textbf{Psychology:} Core courses in psychology that emphasize cognitive processing and research methodology. Any other psychology courses can then be chosen as electives.
        
        
        \textbf{Law:} General law courses with some emphasis on intellectual property rights and negotiations.
        
        
        \textbf{Sports science:} General sports science courses.  
    }%
}

Interestingly, the remaining four official specializations have no corresponding community in our results. This was to be expected, as these remaining specializations are very rarely pursued by students. That is, the communities we have detected are able to represent the specializations that students are actually choosing, but did not reflect other specializations. This is exactly what we expect of community detection, with the added bonus of identifying fields of interests that may not have been previously considered.



\subsection{Analysis of the Typical Student}
To put the community results into perspective, we contrast them with an analysis of the "typical" student's path to graduation. The typical student of the four majors we have focused on (engineering, psychology, business, and computer science) is estimated and visualized using a directed network for each major. In these networks, a node represents a set of courses taken in a particular semester and directed arrows indicate the flow from one semester to the next. The most common semesters are represented as the nodes with the highest weight in the network. As expected, semesters that are clearly laid out in a major’s study plan have the highest weight in the network. However, majors differ greatly in how strictly they define their study plan, and this is also represented in the directed semester networks. Typically, courses are more defined in a program’s first few semesters, but less so as the program progresses and more electives are allowed (see Figures 5 and 6). 

For the major in computer science, this effect of program structure can be seen in an extreme way from the typical student we discovered. This network contained 1408 semester nodes and 1522 edges. The nodes with the most common course combination for each semester reflect the official study plan while also showing the most common combination of elective course selection. For these students, the first two semesters typically consist of mandatory courses only. Out of the 328 graduated students, 174 took the same first semester and 155 the same second semester. However, there is a dramatic drop in the second year in the number of students sharing the most common semester. Of the most common third and fourth semesters course combinations, only 26 and 11 students, respectively, took these most common courses. Similarly, the dispersion of courses increases for the third year, which is also the final year if the official program structure is followed. In the official study plan, the first two semesters have a typical layout of five core courses but no electives while the following semesters all have either two or three electives. These additional electives mean a large increase in the number of possible course combinations for those semesters, which is reflected in the students’ actual choices. Note that under normal circumstances students are allowed to finish any three year program in up to five years, and can also finish in under three years if they have already completed equivalent courses in a different major.

Although the undergraduate program in CS demonstrates this pattern most clearly, all four undergraduate majors we focus on here share the same trend of major structure diversifying as it progresses and more electives are introduced. Despite this, there are differences between majors in the proportion of students belonging to the most common semesters (see Figure 6). These differences are most extreme between psychology and engineering. For psychology, almost all students took the exact same courses for their first semester of the program, while the most common semester of the engineering major was only taken by less than half of the 483 students. As we pooled the different engineering specializations into a single engineering major, this is a reflection of the variety of the specializations within the program. Psychology, however, is the smallest department we are looking at, with less opportunity for diverse course choices. In CS and business, most students share the same courses for the first and second semester. For the CS major, this only lasts for the first two semesters, with a dramatic drop after that in the number of students sharing the most common semester (from 47\% in the second semester to 8\% in the third). For the business major, this drop is more gradual.

\section{Discussion and conclusion}

With this project, we aimed to find whether community detection could be used to effectively identify university students’ fields of interest at NU. To maintain the scope of 
\clearpage
\begin{figure*}
    \centering
    \includegraphics[width=0.75\textwidth]{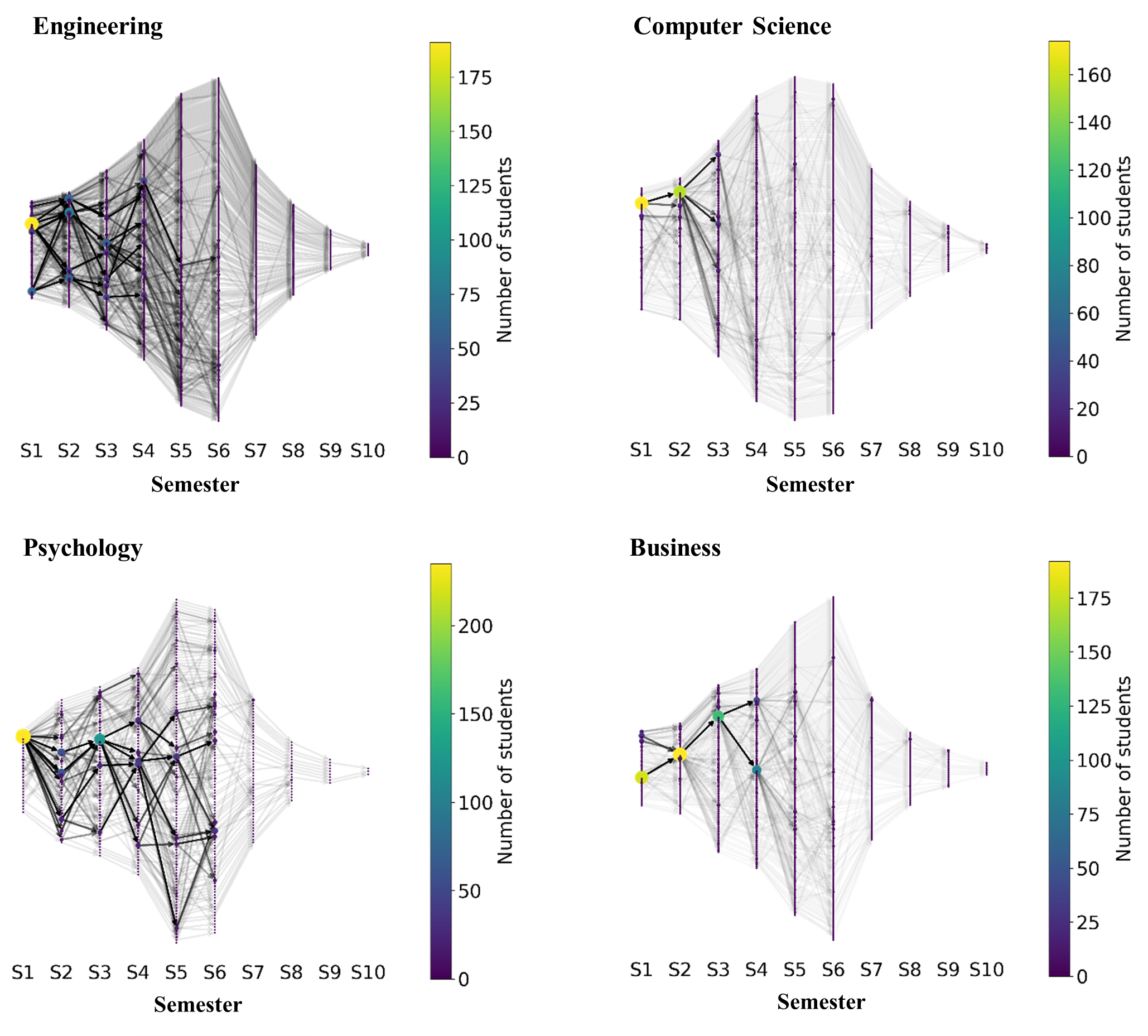}
    \caption{Semester networks representing students in engineering (top left), CS (top right), psychology (bottom left) and business (bottom right). Each node indicates a distinct course combination for a semester. Horizontal lines show all nodes for that semester. The colored bar indicates the number of students who took that specific semester, and the visibility of the edges is determined by how many students went from one semester to the other. }
    \label{fig:my_label}
\end{figure*}

\begin{figure*}
\centering
    \includegraphics[width=0.8\textwidth]{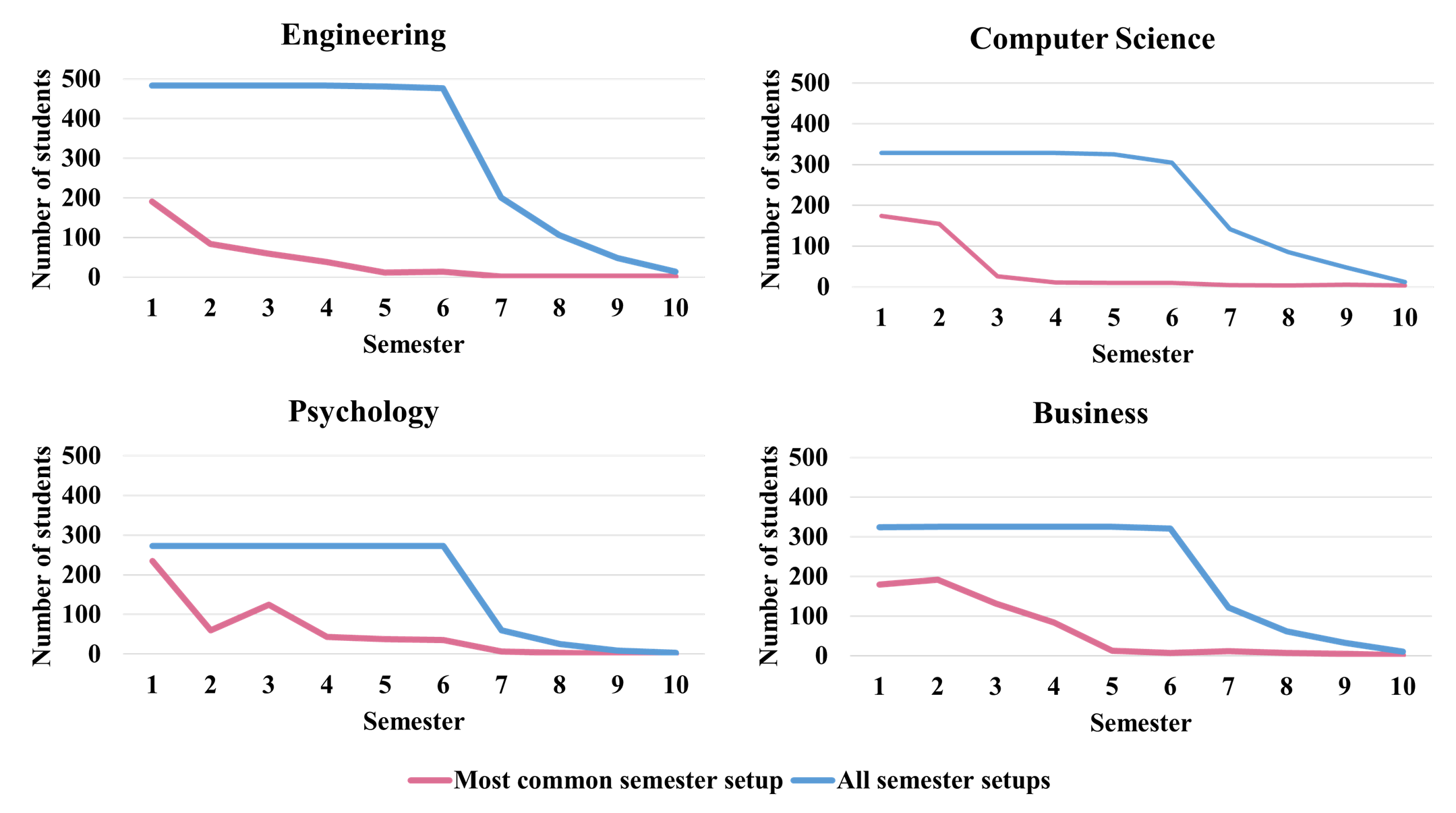}
    \caption{The number of students sharing the most common semester setup throughout the program (red), contrasted by the total number of students in the program as it progresses (blue).}
    
\end{figure*}
\clearpage
result interpretation, we have presented only the results for undergraduate majors in engineering, psychology, business, and computer science. Our resulting communities vary slightly in strength and size, yet almost all of them contain courses of a general theme that seemed to indicate that they are in fact reflecting fields of interest. This builds on the results found by Turnbull and O'Neale. \cite{turnbull_entropy_2020}, who performed community detection on a similar school course network, but without hub removal. This resulted in much more general course communities that demonstrated important but slight differences in the overall majors in their network. When focusing on students’ fields of interests, removing the hubs has allowed us to increase the granularity of the resulting communities while still maintaining community strength and cohesion. However, one of the commonalities between these  

majors is that the largest community detected usually included the major's most popular courses, be that electives or new mandatory courses our hub removal does not consider. As Fortunato \cite{fortunato_community_2016} suggested, using the inter/intra weight density, we were able to evaluate the quality of the communities that were detected with the Louvain algorithm.

The communities we have discovered encapsulate various distinct areas of interests for the different undergraduate majors NU has to offer. Additionally, for the CS department, we have verified that the detected communities also reflect the main areas students choose to specialize in, which further validates our findings. To our knowledge, applying community detection in this way and for this purpose has not been done before. This provides an exciting new tool for universities to better understand their students’ aspirations. 

We also aimed to define the course selection of a typical student of any NU major. We have done this by creating a network where the nodes represent all the semesters taken by each student of that major. We looked at the first 10 semesters as this is the maximum time frame students are given to complete these programs. However, our focus is on the first six semesters which is the time these programs are expected to take when strictly following the study plan. We have found a similar pattern for all majors, where the first semester is the most structured and therefore has the least diverse course choices, with semester course combinations becoming more and more varied as the program progresses. This tells us that perhaps a typical student does not exist in the sense that most students take the same courses throughout their studies, instead the typical student seems to exploit the choices given to them, which is why it may be more informative to focus on specific communities of student's choices. 

The motivation for improving knowledge of student course selection is in part to use this information to provide students with increased flexibility and easier opportunities to pursue their interests and career aspirations. Based on previous studies, we assume that this is the main motivation behind course choices \cite{babad_experimental_2003, maringe_university_2006}. However, these communities may be based on other factors. Examining the characteristics of courses that make up different communities might reveal differences in common factors that determine course selection, such as course difficulty, grading, teacher characteristics, and more \cite{sabot_grade_1991, babad_experimental_2003, maringe_university_2006}. Further, increased flexibility comes with added challenges such as the need for students to better understand the requirements and prerequisites for the courses they want to take, scheduling conflicts, and course difficulty. As students take more control over their studies, data mining can be used to answer these questions and improve student support. For example, course requirements can be identified by extracting information on the learning careers of students who have previously successfully passed the course and comparing it to those who failed the course. Such knowledge can help students adapt to a more liberal course selection process by making more informed choices. This adds to a growing library of data mining methods universities can use to support their students \cite{akpinar_analyzing_2020, yu_towards_2020, zhao_predicting_2020}.

Although we were able to successfully apply network analysis to our student and course data, there were a few setbacks. One drawback in our analysis is the fact that although NU’s administrative data has largely been digitized, this has not always been done in the most structured and data-mining friendly way. For example, all information on specializations (number and names of courses, necessary credits, etc.) was retrieved directly from NU’s website and formatted manually, as this information is not stored in the university's main database. Reliable information on the mandatory courses of each major was also not available, which was why we decided to compare our results when filtering out "mandatory" courses using data-driven hub removal rather than removing the actual mandatory courses. As demonstrated by the high clustering similarity, both methods give comparable results, making hub removal a feasible alternative when this information is missing. Improving data availability, centrality and consistency is currently a priority at NU, but should also be considered by other universities wanting to take full advantage of EDM methods.

Our findings show that network analysis with community detection is a useful tool in understanding students' course selection. The course choice patterns found here can still be explored further. For example, the current results are based on data from students who enrolled in the same program at different times. Thus any small changes in the program structure between years can introduce noise in the data. Looking at individual registration years, perhaps including a larger university with more students, could give clearer results. Further, it would be interesting to repeat the same analysis over separate periods to discover changes in interest fields over time. Finally, it was out of the scope of the current paper to analyze trends based on more detailed characteristics such as gender, age or grades. Augmenting the communities with these factors could provide a tool to identify differences in choices, for example, made by students who graduate successfully and those who struggle more with their studies, perhaps yielding an opportunity for early intervention.  

Educational data mining is an exciting new field with the potential to greatly influence educational institutions and their students going forward \cite{deeva2021introduction}. This project aimed to reveal how network analysis could be used to enhance student course selection by improved understanding of students’ academic interests. Our analysis has successfully led to meaningful results that could easily be replicated by most interested universities with digitized information. Coupling this increased understanding of student interests with added academic support gives universities the tools to raise flexibility within majors while maintaining educational quality. Hopefully, this and other research in the field can be used to offer more tailored and student-led education, which in turns allows students to follow their interests and easily adapt to the ever-changing demands of the job market.


%
\bibliographystyle{abbrv}
\bibliography{sigproc}  

\begin{thebibliography}{10}

\bibitem{akpinar_analyzing_2020}
N.-J. Akpinar, A.~Ramdas, and U.~Acar.
\newblock Analyzing student strategies in blended courses using clickstream
  data.
\newblock In {\em Proceedings of the 13th International Conference on
  Educational Data Mining}, 2020.

\bibitem{babad_experimental_2003}
E.~Babad and A.~Tayeb.
\newblock Experimental analysis of students' course selection.
\newblock {\em British Journal of Educational Psychology}, 73(3):373--393,
  2003.

\bibitem{banerjee_properties_2017}
S.~Banerjee, M.~Jenamani, and D.~K. Pratihar.
\newblock Properties of a projected network of a bipartite network.
\newblock In {\em 2017 International Conference on Communication and Signal
  Processing ({ICCSP})}, pages 0143--0147. {IEEE}, 2017.

\bibitem{barabasi_network_2013}
A.-L. Barabási.
\newblock Network science.
\newblock {\em Philosophical Transactions of the Royal Society A: Mathematical,
  Physical and Engineering Sciences}, 371(1987):20120375, 2013.

\bibitem{ICWSM09154}
M.~Bastian, S.~Heymann, and M.~Jacomy.
\newblock Gephi: An open source software for exploring and manipulating
  networks, 2009.

\bibitem{blansky_spread_2013}
D.~Blansky, C.~Kavanaugh, C.~Boothroyd, B.~Benson, J.~Gallagher, J.~Endress,
  and H.~Sayama.
\newblock Spread of academic success in a high school social network.
\newblock {\em {PLoS} {ONE}}, 8(2):e55944, 2013.

\bibitem{blondel_fast_2008}
V.~D. Blondel, J.-L. Guillaume, R.~Lambiotte, and E.~Lefebvre.
\newblock Fast unfolding of communities in large networks.
\newblock {\em Journal of Statistical Mechanics: Theory and Experiment},
  2008(10):P10008, 2020.

\bibitem{borgatti_network_2009}
S.~P. Borgatti, A.~Mehra, D.~J. Brass, and G.~Labianca.
\newblock Network analysis in the social sciences.
\newblock {\em Science}, 323(5916):892--895, 2009.

\bibitem{cole_intrinsic_2014}
M.~Cole, D.~Bassett, J.~Power, T.~Braver, and S.~Petersen.
\newblock Intrinsic and task-evoked network architectures of the human brain.
\newblock {\em Neuron}, 83(1):238--251, 2014.

\bibitem{deeva2017dropout}
G.~Deeva, J.~De~Smedt, P.~De~Koninck, and J.~De~Weerdt.
\newblock Dropout prediction in moocs: a comparison between process and
  sequence mining.
\newblock In {\em International Conference on Business Process Management},
  pages 243--255. Springer, 2017.

\bibitem{deeva2021introduction}
G.~Deeva, S.~Willermark, A.~S. Islind, and M.~Oskarsdottir.
\newblock Introduction to the minitrack on learning analytics.
\newblock In {\em Proceedings of the 54th Hawaii International Conference on
  System Sciences}, page 1507, 2021.

\bibitem{fortunato_community_2010}
S.~Fortunato.
\newblock Community detection in graphs.
\newblock {\em Physics Reports}, 486(3):75--174, 2010.

\bibitem{fortunato_community_2016}
S.~Fortunato and D.~Hric.
\newblock Community detection in networks: A user guide.
\newblock {\em Physics Reports}, 659:1--44, 2016.

\bibitem{gavrilov_mining_2000}
M.~Gavrilov, D.~Anguelov, P.~Indyk, and R.~Motwani.
\newblock Mining the stock market (extended abstract): which measure is best?
\newblock In {\em Proceedings of the sixth {ACM} {SIGKDD} international
  conference on Knowledge discovery and data mining - {KDD} '00}, pages
  487--496. {ACM} Press, 2000.

\bibitem{gitinabard2018your}
N.~Gitinabard, F.~Khoshnevisan, C.~F. Lynch, and E.~Y. Wang.
\newblock Your actions or your associates? predicting certification and dropout
  in moocs with behavioral and social features.
\newblock {\em International Educational Data Mining Society}, 2018.

\bibitem{Hunter:2007}
J.~D. Hunter.
\newblock Matplotlib: A 2d graphics environment.
\newblock {\em Computing in Science \& Engineering}, 9(3):90--95, 2007.

\bibitem{kardan_prediction_2013}
A.~A. Kardan, H.~Sadeghi, S.~S. Ghidary, and M.~R.~F. Sani.
\newblock Prediction of student course selection in online higher education
  institutes using neural network.
\newblock {\em Computers \& Education}, 65:1--11, 2013.

\bibitem{lancichinetti_community_2009}
A.~Lancichinetti and S.~Fortunato.
\newblock Community detection algorithms: A comparative analysis.
\newblock {\em Physical Review E}, 80(5):056117, 2009.

\bibitem{lin_rotation-invariant_2012}
J.~Lin, R.~Khade, and Y.~Li.
\newblock Rotation-invariant similarity in time series using bag-of-patterns
  representation.
\newblock 39(2):287--315.

\bibitem{lismont_closing_2019}
J.~Lismont, T.~Van~Calster, M.~Óskarsdóttir, S.~vanden Broucke, B.~Baesens,
  W.~Lemahieu, and J.~Vanthienen.
\newblock Closing the gap between experts and novices using
  analytics-as-a-service: An experimental study.
\newblock {\em Business \& Information Systems Engineering}, 61(6):679--693,
  2019.

\bibitem{liu_analyzing_2011}
Y.~Liu, K.~P. Gummadi, B.~Krishnamurthy, and A.~Mislove.
\newblock Analyzing facebook privacy settings: user expectations vs. reality.
\newblock In {\em Proceedings of the 2011 {ACM} {SIGCOMM} conference on
  Internet measurement conference - {IMC} '11}, page~61. {ACM} Press, 2011.

\bibitem{maringe_university_2006}
F.~Maringe.
\newblock University and course choice: Implications for positioning,
  recruitment and marketing.
\newblock {\em International Journal of Educational Management},
  20(6):466--479, 2006.

\bibitem{milliron_exploring_2008}
V.~C. Milliron.
\newblock Exploring millennial student values and societal trends: Accounting
  course selection preferences.
\newblock {\em Issues in Accounting Education}, 23(3):405--419, 2008.

\bibitem{team2014networkx}
{NetworkX developer team}.
\newblock Networkx, 2014.

\bibitem{reback2020pandas}
T.~pandas~development team.
\newblock pandas-dev/pandas: Pandas, Feb. 2020.

\bibitem{que_scalable_2015}
X.~Que, F.~Checconi, F.~Petrini, and J.~A. Gunnels.
\newblock Scalable community detection with the louvain algorithm.
\newblock In {\em 2015 {IEEE} International Parallel and Distributed Processing
  Symposium}, pages 28--37. {IEEE}, 2015.

\bibitem{rienties2018turning}
B.~Rienties and D.~Tempelaar.
\newblock Turning groups inside out: A social network perspective.
\newblock {\em Journal of the Learning Sciences}, 27(4):550--579, 2018.

\bibitem{sabot_grade_1991}
R.~Sabot and J.~Wakeman-Linn.
\newblock Grade inflation and course choice.
\newblock {\em Journal of Economic Perspectives}, 5(1):159--170, 1991.

\bibitem{sarazin_can_2017}
M.~Sarazin.
\newblock Can student interdependence be experienced negatively in collective
  music education programmes? a contextual approach.
\newblock {\em London Review of Education}, 2017.

\bibitem{sarazin_disliking_2021}
M.~A. Sarazin.
\newblock Disliking friends of friends in schools: How positive and negative
  ties can co-occur in large numbers.
\newblock {\em Social Networks}, 64:134--147, 2021.

\bibitem{turnbull_entropy_2020}
S.~M. Turnbull and D.~O'Neale.
\newblock Entropy of co-enrolment networks reveal disparities in high school
  stem participation.
\newblock {\em ArXiv}, abs/2008.13575, 2020.

\bibitem{wellman_network_1983}
B.~Wellman.
\newblock Network analysis: Some basic principles.
\newblock {\em Sociological Theory}, 1:155, 1983.

\bibitem{xu2018many}
Y.~Xu, C.~F. Lynch, and T.~Barnes.
\newblock How many friends can you make in a week?: Evolving social
  relationships in moocs over time.
\newblock {\em International Educational Data Mining Society}, 2018.

\bibitem{yu_towards_2020}
R.~Yu, Q.~Li, C.~Fischer, S.~Doroudi, and D.~Xu.
\newblock Towards accurate and fair prediction of college success: Evaluating
  different sources of student data.
\newblock In {\em 13th International Conference on Educational Data Mining},
  2020.

\bibitem{zhao_predicting_2020}
Y.~Zhao, Q.~Xu, M.~Chen, and G.~Weiss.
\newblock Predicting student performance in a master of data science program
  using admissions data.
\newblock In {\em Proceedings of the 13th International Conference on
  Educational Data Mining}, 2020.

\end{thebibliography}
%
%

\balancecolumns
\end{document}